\documentstyle[twocolumn,pre,aps,psfig]{revtex}
\newcommand{\be}{\begin{equation}}
\newcommand{\ee}{\end{equation}}
\newcommand{\bea}{\begin{eqnarray}}
\newcommand{\eea}{\end{eqnarray}}
\newcommand{\bsa}{\begin{subeqnarray}}
\newcommand{\esa}{\end{subeqnarray}}
\newcommand{\D}{\hbox{D}_x}
\newcommand{\am}{a^{-1}}
\renewcommand{\d}{\hbox{d}_x}
\begin{document}

\title{Kink dynamics in a one-dimensional growing surface}

\author{Paolo Politi$^a$}
\address{Fachbereich Physik, Universit\"at GH Essen, D-45117 Essen (Germany)}
\address{Dipartimento di Fisica dell'Universit\`a di Firenze e Sezione INFM,
L.go~E.~Fermi~2, I-50125 Firenze (Italy)}
\date{\today}
\maketitle

\begin{abstract}
A high-symmetry crystal surface may undergo a kinetic instability
during the growth, such that its late stage evolution resembles a
phase separation process. This parallel is rigorous in one dimension, 
if the conserved surface current is derivable from a
free energy. We study the problem in presence of a physically relevant term
breaking the up-down symmetry of the surface and which can not be
derived from a free energy. Following the treatment introduced
by Kawasaki and Ohta [Physica {\bf 116A}, 573 (1982)] 
for the symmetric case,
we are able to translate the problem of the surface evolution 
into a problem of nonlinear dynamics of kinks (domain walls).
Because of the break of symmetry, two different classes ($A$ and $B$)
of kinks appear and their analytical form is derived.
The effect of the adding term is to shrink a kink $A$ and to widen the
neighbouring kink $B$, in such a way that the product of their widths
keeps constant. Concerning the dynamics, this implies that kinks $A$
move much faster than kinks $B$.
Since the kink profiles approach exponentially the
asymptotical values, the time dependence of the average distance 
$L(t)$ between kinks does {\it not} change: $L(t)\sim\ln t$ in absence of
noise, and $L(t)\sim t^{1/3}$ in presence of (shot) noise. 
However, the cross-over time between the first and the second regime
may increase even of some orders of magnitude.
Finally, our results show that kinks $A$ may be so narrow that their 
width is comparable to the lattice constant: in this case, they indeed
represent a discontinuity of the surface slope, that is an angular point,
and a different approach to coarsening should be used. 
\end{abstract}


\section{Introduction}
\label{introduction}
A crystalline surface growing under a flux of incoming particles from the
vapour phase represents a typical example of an out-of-equilibrium system.
Its microscopic evolution may be described as follows: 
once the adatom has arrived
on the surface, it performs a thermally activated diffusion process till
it is ``trapped" somewhere, or it evaporates by coming back to the
vapour phase. Which surface relaxation mechanism (surface diffusion
or evaporation/condensation) indeed prevails, it depends on the 
temperature and the specific parameters of the material~%
\cite{JVAP}. 
Anyway, for a wide class of materials (mainly metals), at the relevant
temperatures for Molecular Beam Epitaxy (MBE) desorption may be
neglected. In this case, two different ``traps" may be effective:
another adatom (giving rise to a nucleation phenomenon), or a step.
If we limit ourselves to the case of a high-symmetry surface, there are
no preexisting steps and therefore the mentioned step belongs to a
growing island. Once islands have coalesced, leading to the completition
of one layer, steps should disappear and the whole previous process
should start again. 

The previous qualitative picture applies to the case of a stable
layer-by-layer growth. In reality, it is hindered both by noise
and by possible instabilities; sources of noise are fluctuations in the
flux of incoming particles (shot noise), in the surface diffusion
current (diffusion noise), and in the nucleation events (nucleation
noise). Whilst the first two have been well-studied in the context of
several different models~\cite{book_BS}, 
the latter one still needs a more basic comprehension~\cite{noteW}. 
Concerning deterministic instabilities, the main responsible
$-$and perhaps the sole, for a homoepitaxial high-symmetry surface$-$ of the
destabilization of the flat surface, is now known as Ehrlich-Schwoebel
effect~\cite{ES}: an adatom approaching a step from above or below may have
different probabilities of attachement. If the sticking from above is
discouraged, an adatom which has the possibility to choose between 
two different kinds of steps (an ascending one and a descending one)
will stick preferably to the ascending one, thus determining an up-hill
current. It is important to remark that this is a purely 
out-of-equilibrium effect, because at equilibrium detailed balance
forbides such a current! 

Even without entering into details, as it will be done in the next section,
it is possible to explain here the effect of such a mechanism. In fact,
as was firstly pointed out by Villain~\cite{JV}, the resulting surface 
current $j=\nu m$ ($m=\partial z/\partial x$ being the local slope
of the surface and $z$ the local height), once put in the evolution equation: 
$\partial z/\partial t = -\partial j/\partial x$ gives rise to a
diffusion-type equation: $\partial_t z = - \nu \partial^2_x z$,
where the negative sign of the diffusion constant $(-\nu)$ is
responsible for the instability of the flat surface $(z=$~constant). 
In the present paper, we will mainly be concerned with the late stages
of this instability, when additional and nonlinear terms must be
introduced to describe the dynamics of the surface.  
In the next section, we will introduce a more general expression for the
surface current $j$ and we will take into account the
breaking of the $z\to -z$ symmetry, induced by the flux $F$ of atoms.

\section{The surface current}
\label{surface_current}
The study of a growth process may ideally be divided into two main steps:
the first one starts from some microscopic point of view and should 
arrive to a continuum description of the surface/interface; the second one
may even {\it assume} a given evolution equation and study it.
Most of the difficulties encountered in a theoretical study of MBE
are related to the first step. In the present paper, we will limit
ourselves to a one-dimensional high-symmetry surface, and in this section
we will introduce and justify a specific Langevin-type equation.

The local surface height $z(x,t)$ is generally supposed to satisfy a 
{\it local} equation of the form $\partial_t z = aF+{\cal F}(\partial_x z,
\partial^2_x z,\dots)$, $F$ being the incoming flux,
$a$ the in-plane lattice constant, and ${\cal F}$ a 
function of the local profile of the surface (the out-of-plane lattice
constant is put equal to one, i.e. $z$ is adimensional). 
The underlying hypotheses 
have been discussed in Ref.~\cite{PV}, where we have shown that the
appearance of angular points in the surface profile may be treated
correctly solely through the introduction of a nonlocal equation.
We will take up this point again, at the end of the article.

The flux $F$ contains a constant part $F_0$, which is ``eliminated"
by redefining $z(x,t)$: $z\to z - aF_0t$, and a fluctuating part
$\delta F(x,t)$ which represents the so-called shot noise, which is supposed
to follow a Gaussian-like distribution:
\bea
\langle \delta F(x,t)\rangle &=& 0\\
\langle \delta F(x,t)\delta F(x',t')\rangle &=& 2F_0 \delta(x-x')
\delta(t-t')
\eea

In the limit of negligible desorption, and if overhangs are forbidden,
surface growth proceeds by conserving both mass and volume: therefore,
the function ${\cal F}$ must be derivable from a surface 
current $j$, and the evolution equation will be written in the form:
\be
\partial_t z(x,t) = - a \partial_x j + a\delta F(x,t)~,
\label{lan_eq}
\ee

The central question is which current $j$ governs the evolution of the
surface: a still debated question even for the simplified model of
a one dimensional surface, as shown by the following discussion on the
different terms appearing in $j$.
Symmetry arguments simply tell that $j$
does not depend on $z$~\cite{z_inv}, but on its derivatives ($m=\partial_x z,
m'=\partial^2_x z, m''=\partial^3_x z,\dots$) 
and that $-$on a high-symmetry surface$-$
it must be an odd function of $x$: so, a term proportional to
$m$ or $m''$ satisfies this request, but if proportional to $m'$, it does not.

\subsection{Ehrlich-Schwoebel current}
In the Introduction, we mentioned the Ehrlich-Schwoebel effect,
which gives rise to a slope-dependent current: $j_{ES}(m)$.
Since it must be an odd function of $m$, its form at small $m$ will be
$j_{ES}=\nu m$. The coefficient $\nu$ depends~\cite{P,review_Krug}
on the flux $F_0$, the diffusion length $\ell_D$ and the Schwoebel
length $\ell_S$: $\ell_D$ measures the typical linear distance travelled by
the adatom before meeting another one and forming the nucleus of a
growing island. It represents the ``maximal" size of a terrace, because
if $\ell > \ell_D$ the probability to nucleate a new island on it is
very high; during the first stages of growth, when the surface is still
more or less flat, $\ell_D$ is also the typical size of a terrace.
It is not so when an instability develops: in this case, $\ell$ may be
much smaller than $\ell_D$ and the slope $m=1/\ell$ may be fairly large. 
Indeed, the slope $1/\ell_D$ discriminates between a 
nucleation-dominated regime ($m\ll 1/\ell_D$) and a step-flow regime
($m\gg 1/\ell_D$): the latter is generally relevant for vicinal
surfaces, which grow through sticking of adatoms to preexisting steps;
anyway, if a flat surface develops an instability with regions of high
slope, such regime becomes important also for high-symmetry orientations.

The second relevant length, the Schwoebel length, is a measure of the
asymmetry in the sticking coefficients of an adatom to a step. Its
simplest form~\cite{JVAP} is: $\ell_S=a(D_-/D_+-1)$, where $D_+$ and $D_-$
are such coefficients for an adatom approaching the step from
above ($D_+$) and below ($D_-$). The existence of an Ehrlich-Schwoebel
effect means that $D_+<D_-$ and therefore $\ell_S >0$. To describe the
meaning of $\ell_S$, let us consider a terrace of size $\ell$~($<\ell_D$):
if $\ell_S\ll\ell$, only a fraction $\ell_S/\ell$ of the fallen adatoms
will contribute to the uphill current, and therefore 
$j_{ES}=(F_0\ell)(\ell_S/\ell)=F_0\ell_S$,
since the number of atoms arriving per unit time on a terrace of size
$\ell$ is nothing but $F_0\ell$. Conversely, if $\ell_S\gg\ell$ all
the adatoms will stick to the ascending step, and so:
$j_{ES}=F_0\ell$. The simplest interpolation formula, valid for any value
of $\ell_S$ is:
\be
j_{ES} = {F_0\ell_S\over 1 +\ell_S|m|}~~~~~~|m|>1/\ell_D
\label{j_s2}
\ee
This formula also allows to obtain a semiquantitative expression for the
parameter $\nu$: in fact, when $|m|\simeq 1/\ell_D$, Eq.~(\ref{j_s2})
must match the expression valid at small slopes: $j_{ES}=\nu m$.
The result is: $\nu=F_0\ell_S\ell_D^2/(\ell_S+\ell_D)$.
It is important to remark that all the previous considerations may be
made more rigorous~\cite{PV}, but in this section we are mainly 
interested in justifying the expression for the current $j$, 
rather than in deriving it.

The main characteristic of the Ehrlich-Schwoebel current just discussed 
is that it has no zeros other than $m=0$ and $m=\pm\infty$. A zero in
$j_{ES}$ is extremely important~\cite{KPS}, 
because the other terms in $j$ will be
seen to depend on higher order derivatives of $z(x,t)$. So, a constant
slope $m_0$ may be a stationary slope if and only if $j_{ES}(m_0)=0$.
An extra-zero $m_0$ may have different origins: the symmetry of the
crystal lattice~\cite{KPS,AF}, nonthermal relaxation 
mechanisms~\cite{non_thermal}, or a transient mobility of the adatom just
after the deposition~\cite{SV}.
For example, the slope at 45 degrees corresponds in a cubic lattice to
the high-symmetry orientation (11): we expect that $j_{ES}$ vanishes
on it, as it vanishes on the (10) ($m=0$) and (01) ($m=\infty$)
orientations. A different example is the following: if atoms falling 
in the vicinity of a step have a higher probability to land on the
lower terrace, or to kick down the step adatom, a down-hill current 
$j_\downarrow$,
proportional to the density of steps and therefore to the slope $m$,
will appear: $j_\downarrow=-\nu'm$. So, if $\nu>\nu'$, a zero will
appear when $(j_{ES}+j_\downarrow)=0$.

Whatever is the origin of extra-zero(s) in the slope-dependent current,
we can introduce two different models, according to the presence (model~I)
or absence (model~II) of zeros at finite slope. The simplest expressions
of $j_{ES}$ for the two models, having the correct symmetry
properties are~\cite{note1}:
\bea
\hbox{model I\phantom{I}~~~~~~~~~}j_{ES} &=& \nu m (1-m^2/m_0^2)\label{j_sI}\\
\hbox{model II~~~~~~~~~}j_{ES} &=& {\nu m \over 1 +\ell_D^2 m^2}
\label{j_sII}
\eea
Model II does not correspond to a phase separation process 
(see Sec.~\ref{standard_model}):
it will be discussed in Sec.~\ref{Sec_disc}.

\subsection{Mullins-like current}

The most famous ``equilibrium" current is perhaps the one ($j_M$) introduced by
Mullins~\cite{Mullins} forty years ago, to study the relaxation towards
equilibrium of a non-singular grooved surface.
A simple derivation starts from writing $j_M$ as the gradient of a
chemical (surface) potential: $j_M = -\Gamma\partial_x\mu$, where
$\Gamma$ is the adatom mobility, and afterwards to derive $\mu$ from
a surface free-energy:
\be
\mu = {\delta{\cal E}\over\delta z(x)},~~~~~~~\hbox{with}~~~
{\cal E}=\sigma\int dx\sqrt{1+m^2a^2}
\ee
By combining the different equations, in the limit of small slopes we obtain:
\be
j_M = K m''(x) \label{jk}
\ee
with $K=a^2\Gamma\sigma$. 

The usage of this expression in our problem may be questionable in
at least two respects: first, it applies to a nonsingular surface, i.e.
above the roughening transition $T_R$; second, it applies to a close-to%
-equilibrium surface. Concerning the first remark, our surface is
a high-symmetry one and therefore almost necessarily below $T_R$,
because for a high symmetry orientation the roughening temperature
is equal or nearly equal to the melting temperature $T_M$, while
ordinary temperatures for MBE are well below $T_M$. 
Nevertheless, our surface $-$which is strongly out-of-equilibrium$-$
contains a lot of steps because the incoming flux makes the surface 
rough~\cite{Rodi}: therefore, the surface current should be
nonsingular at zero slope. 

The latter remark
is more ``critical": the Mullins current derives from thermal
detachment of atoms from steps, in order to minimize the surface
free-energy. It is not clear if such process is effective in
presence of a flux $F$. For example, Stroscio and Pierce~\cite{SP}
state that thermal detachment is negligible in the homoepitaxial
growth of Fe (at least at room temperature) and therefore they do not
write~\cite{Stroscio} such a term in the current.
Anyway, it has been shown~\cite{PV,Rodi}
that the current (\ref{jk}) may derive also from nonequilibrium
effects: nucleation noise and diffusion noise. The first one should
be dominant and correspond to the value~\cite{PV,Rodi}:
$K=F_0\ell_D^4$.

\subsection{Symmetry-breaking current}

The terms in the surface current which have been introduced so far 
not only satisfy the $x\to -x$ symmetry (because $j(-x)=-j(x)$),
but they also fulfil the up-down symmetry, corresponding to the
change of sign of $z$. In fact, if $z\to -z$ both $j_{ES}$ and
$j_M$ change sign. However, there is no reason to expect that
surface growth proceeds by conserving such symmetry, since the
flux breaks it. 

A symmetry-breaking (SB) term is intrinsically nonlinear, because
any current of the form $j\sim\partial^n_x z(x,t)$ changes sign
with $z$. The lowest order expression which changes sign with $x$
but does not change sign with $z$ is:
\be
j_{SB} = \partial_x A(m^2)
\label{9}
\ee
where $A$ is any even function of the local slope. 
The simplest form for $A$: $A=(\lambda/2)m^2$ has been introduced 
by Sun et al.~\cite{Sun}.
It is also called ``conserved Kardar-Parisi-Zhang term", because in
Eq.~(\ref{lan_eq}) it looks like the laplacian of $(\partial_x z)^2$,
i.e. the nonlinear term of the KPZ equation~\cite{KPZ}.

The current (\ref{9}) is not derivable from a free energy.
As pointed out by Somfai and Sander~\cite{SS} it is necessary to
rise the order of $j_{SB}$ to make it derivable from some free energy
(for example, $j_{SB}\sim\partial_x[(m')^2]={\delta\over\delta m}
\int dx {\cal F}_{SB}$ with ${\cal F}_{SB}\sim (m')^3$\,).

Before proceeding, let us discuss the physical
origin of $j_{SB}$. When there is a gradient in the density $\rho$
of adatoms, a current of the form $j=-D \partial_x\rho$ is
expected, where $D$ is the diffusion constant.
In the case of a growing surface, the
applicability of the previous expression is not obvious, because steps
are sink for diffusing atoms and $-$at least if thermal detachment is
forbidden$-$ interlayer diffusion is absent. In spite of this, the above
expression may help in understanding: in fact, adatom density on a
terrace depends on its size $\ell$, because a larger terrace collects
more atoms from the flux than a smaller one. So, $\rho=\rho(\ell)=
\rho(|m|)$. In other words, the function $A$ appearing in $j_{SB}$ 
seems to be proportional to the adatom density itself.

This interpretation can be made more rigorous for large slopes
$(|m|=1/\ell > 1/\ell_D)$, where nucleation of new terraces is absent
and $\rho$ can be simply determined by solving the diffusion
equation $\partial_t\rho=F_0+D\partial_x^2\rho$ in the
quasi-static approximation $(\partial_t\rho=0)$ and with $\rho(0)=
\rho(\ell)=0$ as boundary conditions (i.e. steps are perfect sinks).
The resulting average density on the terrace is $\rho\simeq (F_0/D)
\ell^2$ and the current is $j_{SB}\simeq -F_0\partial_x (1/m^2)$.
This expression agrees with those determined, with different methods, by
Politi and Villain~\cite{PV} and by Krug~\cite{krug2}.
Hunt et al.~\cite{Hunt} suggest that $j_{SB}$ may derive
from the sticking-asymmetry induced by the Ehrlich-Schwoebel effect:
nevertheless, $j_{SB}$ does not vanish even if $\ell_S=0$~\cite{PV,krug2}.

One could ask why the average value of $\rho$ is taken. The answer is that
inhomogeneities in the adatom density on a given terrace give rise to
$j_{ES}$! In fact, if no Ehrlich-Schwoebel effect is present, $\rho(x)$
is symmetric with respect to the center of the terrace and therefore
the average value of $\partial_x\rho$ vanishes. Conversely, if $\ell_S>0$
then $\langle\partial_x\rho\rangle_{\hbox{terrace}}\ne 0$ and it corresponds
just to $j_{ES}$. This remark stresses the ``similar" origin of $j_{ES}$
and $j_{SB}$. It is likely that a systematic derivation of the surface current
should give all the terms we have introduced: 
$j_{ES}$ (which depends on the slope $m$), $j_{SB}$ (which depends on the
curvature $m'$), and $j_M$ (which depends on a higher order derivative:
$m''$).
Anyway, a rigorous derivation is still lacking at the moment,
above all for a high-symmetry orientation.

\subsection{The current of our model}
In the following, we will study the dynamical evolution of the surface,
as determined by the current: 
\be
j=j_{ES}+j_M+j_{SB}
\label{current}
\ee
where:
\bea
j_{ES} &=& \nu m\left(1-{m^2\over m_0^2}\right)\\
j_M    &=& Km''\\
j_{SB} &=& \lambda m m'
\eea

The reason of our choice is clear: we want to study the effect of the
symmetry-breaking current ($j_{SB}$) on the phase separation process determined
by the other two terms of the surface current ($j_{ES}+j_M$), 
and at this aim we choose
the simplest expression for $j_{ES}$ $-$which must have a zero at a finite
slope $m_0-$ and for $j_{SB}$ $-$for which we take $A(m^2)=\lambda m^2/2$.
In the last section, we will discuss how the conclusions depend or
{\it not} depend on the present choice.

\section{Evolution in absence of the symmetry-breaking current}
\label{standard_model}
In the ``language" of surface growth, the evolution of the surface
proceeds as follows: after a time $t^*$ an instability of the flat
surface with a well-determined wavelength $L^*$ develops. In this linear
regime, $L^*$ is constant and the amplitude increases exponentially.
Afterwards, because of the nonlinearity of $j_{ES}$ a coarsening process
takes place: the wavelength $L(t)$ of the mound-like (or pyramid-like)
surface profile increases in time, whilst the maximal slope tends to the
constant values $\pm m_0$. So, the surface is ``made up" of neighbouring
regions where the slope is alternately (nearly) equal to $+m_0$ and
$-m_0$. 

The first stages of growth can be analyzed by linearizing Eq.~(\ref{lan_eq})
with the current (\ref{current}):
\be
a^{-1}\partial_t z(x,t) = -\nu\partial^2_x z(x,t) - K\partial^4_x z(x,t)
\ee 
which shows~\cite{PV,review_Krug,Hunt} 
that the flat surface is unstable against
deformations of wavelength larger than $L_c=2\pi\sqrt{K/\nu}$.
The most unstable mode correspondes to $L_u=\sqrt{2}L_c$ and its
amplitude grows as $\exp\left[(a\nu^2/4K)t\right]$. So, $L^*=L_u$
and $t^*=(4K/a\nu^2)$.

The nonlinear profiles of the mounds are determined as stationary
solutions of Eq.~(\ref{lan_eq}), that is to say as solutions of the
equation $j=0$:
\be
j_{ES}(m) + Km''(x) =0
\ee
This equation can be derived by the following Lagrangian:
$$
{\cal L} = (K/2)m'^2 - V(m)~~~~~~~~\hbox{with }~V'(m)=j_{ES}(m)
$$
which corresponds to an anharmonic pendulum, once we have identified the
slope $m$ as its spatial coordinate and $x$ as the time.
Since the potential $V(m)=(\nu/2)m^2[1 - m^2/2m_0^2]$ 
has two symmetric maxima in $\pm m_0$, the
period of the oscillation (i.e. the wavelength of the surface profile)
diverges when its amplitude (i.e. the maximal slope of the surface profile)
goes to $m_0$. If $j_{ES}$ followed model~II, $V(m)$ would have no
maxima and no limitation on the slope would be present. 

By going on with this mechanical analogy,
the existence of coarsening requires a condition on the stationary 
configurations: the period of the oscillation must be an increasing
function of the amplitude~\cite{opl}; a condition which is surely fulfilled by
the potential $V(m)$, since the quartic correction has a negative sign!
Clearly, coarsening also requires that these stationary solutions are
not stable: more precisely, they must be unstable with respect
to wavelength fluctuations, but stable with respect to amplitude
fluctuations.

The previous mechanical analogy helps in understanding why the surface
keeps a regular profile and also allows to determine this profile at
a given time, but it is not effective in determining the time
dependence of $L(t)$, i.e. the coarsening law~\cite{Hunt}. 
To this end, we must
observe that the evolution equation for the local slope $m$ (which
represents the ``order parameter" of our problem) satisfies the noisy
Cahn-Hilliard equation~\cite{review_Bray}:
$$
\am\partial_t m = \partial^2_x \left({\delta {\cal F}\over\delta m}\right)
+\eta(x,t)~~~~~~~~\hbox{where~~}{\cal F}=\int\! dx {\cal L}
$$

This equation corresponds to a phase separation process, where the
order parameter is conserved ($\partial_t\int\! dx\, m(x,t) =0$). The system
is made up of domains where $m$ equals one of the two degenerate minima
of the potential energy $U(m)=-V(m)$; domains which are separated by
domain walls move in order to minimize the ``action" ${\cal F}$.
Domain wall (or ``kink") movement is determined both by their
(deterministic) interaction and by fluctuations induced by the conserved
noise. We will see that the growing surface (even in presence of the 
symmetry-breaking current $j_{SB}$) can be mapped in a one-dimensional
system of interacting kinks which annihilate,
so that the average distance $L(t)$ between kinks increases in time.

By using this method for the symmetric case ($j_{SB}=0$), 
Kawasaki and Ohta~\cite{KO} have found the
equation of motion for the kinks, which has been then studied by
Kawakatsu and Munakata~\cite{KM}. The final result is that $L(t)$
grows logarithmically with time if noise is absent and grows as $t^{1/3}$
if noise is present. 

\section{Kink profiles}
\label{kink_profiles}

A stationary kink $M(x)$ is defined as a monotonic solution of $j[M(x)]=0$,
with $M(x)$ tending to (different) minima of $U(m)$, when $x\to\pm\infty$.
In the present case, there are only two symmetric minima in $\pm m_0$ and
therefore only two kinks $M_\pm (x)$ are possible, the subscript 
corresponding to the sign of its first derivative, i.e. to the curvature
of the surface profile.

The surprising result is that the ``shape" of the kink does {\it not}
change because of the introduction of the symmetry-breaking term.
To see it, let us replace the expression:
\be
M_\pm(x) = \pm m_0\tanh(\kappa_\pm x/2)
\label{kink_pro}
\ee
in the differential equation $j=0$:
\be
Km''(x) + \nu m(1-m^2/m_0^2) + \lambda m m' = 0
\label{sta_eq}
\ee

We obtain the following second degree equation for the parameters
$\kappa_\pm$:
\be
K\kappa_\pm^2 \mp\lambda m_0\kappa_\pm -2\nu =0
\ee
which gives the positive
solutions:
\be
\kappa_\pm = \left( \sqrt{\lambda^2 m_0^2 + 8\nu K} \pm \lambda m_0\right)
/2K
\label{kappa_pm}
\ee

Two limiting cases, corresponding to weak and strong symmetry-breaking,
will be frequently used:
\bea
\lambda m_0 \ll \sqrt{8\nu K}~~~~~~~&& \kappa_+=\kappa_-=\sqrt{2\nu/K}
\equiv\kappa_0\\
\lambda m_0 \gg \sqrt{8\nu K}~~~~~~~&& \kappa_+=\lambda m_0/K~~
\kappa_-=2\nu/\lambda m_0
\label{kappa}
\eea
So, the effect of $j_{SB}$ is two create two classes of kinks:
kinks ``$A$", given by the profile $M_+(x)$ and characterized by a 
width $(1/\kappa_+)$, and kinks ``$B$", given by the profile $M_-(x)$ and
whose width is $(1/\kappa_-)$. For a strong $j_{SB}$, $\kappa_+\gg\kappa_-$:
kinks $A$ are much narrower than kinks $B$. It must also be observed that
the product ($\kappa_+\kappa_-$) does not depend on $\lambda$, since it equals
(see the algebraic equation) $(2\nu/K)$. In other terms, the effect of $j_{SB}$
is to shrink kinks $A$ and to widen kinks $B$, in such a way that the product
of their widths keeps constant.

\section{From surface dynamics to kink dynamics}

In this section we will describe the method to solve the growth equation for 
the surface-slope profile:
\be
\am\partial_t m = \D [Km'' - U'(m) +\lambda mm']
~~~~\D=-\partial^2_x 
\label{1}
\ee
in a ``multi-kink" approximation. Since our approach follows that one
introduced by Kawasaki and Ohta~\cite{KO} to study the above equation
in absence of the $\lambda$-term, we will expose the main calculations
in App.~\ref{app_A} and here we will limit ourselves 
to explain the general lines of the method.

Once a kink is inserted in our problem, it moves with a given (constant)
velocity $v^0$ and a profile $m(x,t)=M(x-v^0 t)$, where $v^0$ is found by
solving the eigenvalues problem obtained by putting $m(x,t)$ in (\ref{1}).
Our system is made up of an ensemble of kinks $A$ which alternate to
kinks $B$, and we will look for an approximate solution of (\ref{1}) as
a superposition of kinks centered in $x_i$ and moving with velocity $v_i$.
Because of the interaction between kinks, $v_i$ is not a constant, and
depends on the position of the other kinks. In principle, the nonlinear
part of $U(m)$ (i.e. the quartic term $m^4$) gives rise to terms of
$n$-kinks interaction: we will adopt a ``binary-interaction" approximation,
which will be further simplified by limiting to nearest-neighbour interaction.
This procedure is justified by the fact that we are interested in the late
stages of growth, when the distance between kinks is much larger than the
width of their cores ($=1/\kappa_\pm$): so, they interact only through
the tails of the profiles, which means that the interaction decays
exponentially, since $\tanh(\kappa x/2)\simeq \pm 1 \mp \exp(-\kappa |x|)$
when $x\to\pm\infty$. For the same reason, the velocities $v_i$ and the
accelerations $\dot v_i$ will be considered ``small", because
the typical size of the
mounds grows slower than linearly: This means that the velocity of the
coarsening process goes to zero, as time increases.

As a final result, we obtain a Langevin equation for the discrete
variables $x_i(t)$, or $-$equivalently$-$ for the kink-kink distances
$X_i(t)\equiv x_{i+1}(t)-x_i(t)$, which will be studied by translating
it in a Fokker-Planck equation.

The treatment of Eq.~(\ref{1}) (see App.~\ref{app_A}) gives the following 
coupled equations for the kink positions:
\bea
&&-2\am m_0^2\sum_j (-1)^{i-j}|x_i-x_j|\dot x_j =\nonumber\\ 
&& (C_1)+(C_2)+(C_4) + \eta_i(t)
\label{geppetto}
\eea
where:
\bea
(C_1) &=& 8\nu m_0^2 [R_\beta(X_i)-R_\beta(X_{i-1})]\nonumber\\
(C_2) &=& \beta (4/3)m_0^3\kappa_\beta\lambda
[R_{-\beta}(X_i)-R_{-\beta}(X_{i-1})]\label{Ci}\\
(C_4) &=& -\beta 4m_0^3\kappa_\beta\lambda [R_\beta(X_i)-R_\beta(X_{i-1})]
\nonumber
\eea
and:
\bea
\langle\eta_i(t)\rangle&=&0\nonumber\\
\langle\eta_i(t)\eta_j(t')\rangle&=&-4m_0^2 F_0(-1)^{i-j}|x_i-x_j|\delta(t-t')
\nonumber
\eea

Let us explain the notations: The $i$-th kink
is centered in $x_i$, and $-$because of the breaking of symmetry$-$ 
two different classes of kinks exist. 
In accordance with Sec.~\ref{kink_profiles},
their profiles are given by: $M_\beta(x)=\beta m_0\tanh(\kappa_\beta x/2)$,
where $\beta=\pm 1$. We will assume that the $i$-th kink is of
class $\beta$ (whatever is its value) and its nearest-neighbours of class
$-\beta$. The quantity:
\be
R_\beta(x) = \exp(-\kappa_\beta x)
\ee
in the ($C_i$) expresses the interaction between kinks, when the
distances $|x_{i\pm 1}-x_i|$ are large compared to $(1/\kappa_\beta)$.

Eq.~(\ref{geppetto}) can also be written
in matrix form: $\hbox{\boldmath $A$}_{ij}\dot x_j = I_i +\eta_i$.
The matrix $\hbox{\boldmath $A$}$ takes into account the kinematical
coupling between kinks, due to the conservation of the order parameter,
and $\underline{I}$ contains the forces between kinks. The matrix
$\hbox{\boldmath $A$}$ can be inverted~\cite{KM}, giving a
tridiagonal and symmetric $\hbox{\boldmath $A$}^{-1}$:
\bea
\hbox{\boldmath $A$}^{-1}_{ii} &=& {a\over 4m_0^2}\left({1\over X_i}+
{1\over X_{i-1}}\right)\\
\hbox{\boldmath $A$}^{-1}_{i+1,i} &=& {a\over 4m_0^2}{1\over X_i}
\eea
The evaluation of $\hbox{\boldmath $A$}^{-1}\underline{I}$ is trivial:
\be
(\hbox{\boldmath $A$}^{-1}\underline{I})_i =
{a\over 4m_0^2}\left({I_i+I_{i+1}\over X_i}+{I_i+I_{i-1}\over X_{i-1}}\right)
\ee
and the explicit expression of $I_i$ is found directly from Eqs.~(\ref{Ci}):
\be
I_i = R_\beta^*(X_i)-R_\beta^*(X_{i-1})
\ee
where $R_\beta^*(X)$ is a linear combination of the two different
$R_\beta(X)$:
\bea
&&R_\beta^*(X)\equiv c_\beta R_\beta(X) + d_\beta R_{-\beta}(X)~~~~~
\hbox{with}\nonumber\\
&&c_\beta=8\nu m_0^2-\beta4m_0^3\kappa_\beta\lambda~~~~~~\hbox{and}~~~~~~
d_\beta=\beta(4/3)m_0^3\kappa_\beta\lambda\nonumber
\eea

Concerning the noise, it is preferable to work with quantities which are
not spatially correlated. To this end, the matrix $\hbox{\boldmath $A$}^{-1}$
is written as the product $\hbox{\boldmath $P$}\hbox{\boldmath $P$}^T$
and new noise variables $\underline{\tilde\eta}=\hbox{\boldmath $P$}^T
\underline{\eta}$ are defined. Since $\hbox{\boldmath $P$}$ is a
bidiagonal matrix whose nonvanishing elements are:
\be
\hbox{\boldmath $P$}_{ii}=\hbox{\boldmath $P$}_{i+1,i}=
{\sqrt{a}\over 2m_0}{1\over\sqrt{X_i}}
\ee
$\tilde\eta_i$ is given by $\tilde\eta_i=\sqrt{a}
(\eta_i+\eta_{i+1})/(2m_0\sqrt{X_i})$, and it results that:
\be
\langle\tilde\eta_i(t)\rangle=0~~~~~
\langle\tilde\eta_i(t)\tilde\eta_j(t')\rangle = 2aF_0\delta_{ij}\delta(t-t')
\ee
In order to eliminate the constant factor in the correlator, we simply put:
$\tilde\eta_i=\sqrt{2aF_0}\xi_i$. This way, the final equation for kink
dynamics is:
\bea
\dot x_i(t)&=&{a\over 4m_0^2}\left[{I_i+I_{i+1}\over X_i}+
{I_i+I_{i-1}\over X_{i-1}}\right]\nonumber\\
&+&{\sqrt{2F_0}a\over 2m_0}\left[{\xi_i\over\sqrt{X_i}} + 
{\xi_{i-1}\over\sqrt{X_{i-1}}} \right]\label{v_i}\\
\langle\xi_i(t)\xi_j(t')\rangle&=&\delta_{ij}\delta(t-t')\nonumber
\eea

\section{Forces and kink velocities}
In this section we want to discuss the effect of the
symmetry breaking on the equations of motion for the kinks.
Since we are here interested in the deterministic part of the interaction,
we will not consider the noise. Therefore
Eq.~(\ref{geppetto}) takes the form: $\hbox{\boldmath $A$}_{ij}\dot x_j = I_i$.
Kawasaki and Ohta~\cite{KO} suggest to look on $I_i$ as the force 
acting on the $i$-th kink. Let us consider the two opposite limits:
$\lambda=0$ and $\lambda m_0\gg\sqrt{8\nu K}$. For $R^*_\beta(X)$ we obtain:
$$
R^*_\beta(x) = 8\nu m_0^2\exp(-\kappa_0 x)
$$
in the first limit ($\lambda=0$), and:
\bea
R^*_+(x) &=& (4\lambda^2 m_0^4/3K)\exp(-\kappa_\lambda x)\nonumber\\
R^*_-(x) &=& 16\nu m_0^2\exp(-\kappa_\lambda x)\nonumber
\eea
in the second one ($\lambda m_0\gg\sqrt{8\nu K}$). In the previous
equations, $\kappa_0=\sqrt{2\nu/K}$ and $\kappa_\lambda=(\lambda m_0/K)
\gg\kappa_0$.
They correspond to $\kappa_-$ in the two pertinent limits.

In the case of absence of the $\lambda$-term, we simply get:
\be
I_i = 8\nu m_0^2[\exp(-\kappa_0 X_i)-\exp(-\kappa_0 X_{i-1})]
\label{I0}
\ee
This equation can be interpreted by saying that there is an attraction
between kinks, proportional to $\exp(-\kappa_0 X)$.
If $\lambda\ne 0$ (and ``strong"), than we must distinguish between positive
and negative kinks:
\bea
I_i &=& (4\lambda^2 m_0^4/3K)[\exp(-\kappa_\lambda X_i)-
\exp(-\kappa_\lambda X_{i-1})]~(\beta>0)\nonumber\\
I_i &=& 16\nu m_0^2[\exp(-\kappa_\lambda X_i)-\exp(-\kappa_\lambda X_{i-1})]
~(\beta<0)\nonumber
\eea
The first comment is that symmetry breaking implies that a positive kink
is attracted (by a negative one) more strongly than a negative kink is
attracted by a positive one. In other words, if we assign a mass to a kink,
a negative kink weights more than a positive one, and the mass is
proportional to the width of the kink itself.

This interpretation seems to be satisfactory, but if we analyse the
velocities $\dot x_i(t)$ rather than the ``forces" $I_i$ the picture
becomes more complicated.
In the limit $\lambda=0$ we have:
\bea
\dot x_i(t) = 2a\nu&&\left[{\exp(-\kappa_0 X_i)\over X_{i-1}} -
{\exp(-\kappa_0 X_{i-1})\over X_{i}}\right.\nonumber\\
&+& \left.{\exp(-\kappa_0 X_{i+1})\over X_{i}} -
{\exp(-\kappa_0 X_{i-2})\over X_{i-1}} \right]
\label{v0}
\eea
So, the effect of the conservation law (i.d. of the matrix
$\hbox{\boldmath $A$}$) is that $\dot x_i(t)$ depends not only on the positions
of the nearest neighbour (nn) kinks $(x_{i\pm 1})$, but also on those of the
next-nearest (nnn) ones $(x_{i\pm 2})$. Even more important, the nnn-%
``interaction" is of the same order of magnitude as the nn-one! 
Whilst the interpretation of $\exp(-\kappa_0 X_i)$ and $\exp(-\kappa_0 X_{i-1})$
in Eq.~(\ref{I0}), as $-$respectively$-$ the interaction with the kinks $(i+1)$
and $(i-1)$ is straightforward, in Eq.~(\ref{v0}) the generic term
$\exp(-\kappa_0 X_l)$ is divided by a different $X_j$, and therefore a
similar interpretation becomes less evident. Anyway, if we don't ascribe too
much importance to the quantities $X_j$ in the denominator, Eq.~(\ref{v0}) says
that kink $i$ is attracted both by nn-kinks and nnn-kinks: the ``interaction"
between $i$ and $(i\pm 2)$ has a kinematical origin (conservation of the order 
parameter) and indeed depends on $(x_{i\pm 2}-x_{i\pm 1})$ rather than on
$(x_{i\pm 2}-x_i)$. A further comment is that in the evaluation of $(I_i +
I_{i+1})$ two terms cancel exactly, because in this case action and
reaction are opposite and equal. 

If now we consider the case of a strong symmetry breaking term, the velocity
takes the form:
\bea
\dot x_i(t)|_{\beta>0} &=& {a\lambda^2 m_0^2\over 3K}\left[
\left({1\over X_i}+{1\over X_{i-1}}
\right)\exp(-\kappa_\lambda X_i)\right.\nonumber\\
&&-\left. \left({1\over X_i}+
{1\over X_{i-1}}\right)\exp(-\kappa_\lambda X_{i-1})\right]\nonumber\\
&& +4a\nu\left[ {\exp(-\kappa_\lambda X_{i+1})\over X_i}
-{\exp(-\kappa_\lambda X_{i-2})\over X_{i-1}}\label{vl+}\right]\nonumber
\eea
and:
\bea
\dot x_i(t)|_{\beta<0} &=& 
-{a\lambda^2 m_0^2\over 3K}\left[{\exp(-\kappa_\lambda X_i)\over X_i}
-{\exp(-\kappa_\lambda X_{i-1})\over X_{i-1}}\right.\nonumber\\
&& +\left.{\exp(-\kappa_\lambda X_{i+1})\over X_i}
-{\exp(-\kappa_\lambda X_{i-2})\over X_{i-1}}\right]
\label{vl-}
\eea
The surprising result is that the sign of the terms proportional to
$\exp(-\kappa_\lambda X_i)$ and $\exp(-\kappa_\lambda X_{i-1})$
is inverted: so
$-$because of kinematics$-$ a negative kink is
subject to a repulsive interaction with its nn-kinks! This result derives
from the unbalancing of action and reaction. A closer inspection of the 
derivation of Eqs.~(\ref{vl+}-\ref{vl-}) allows to give the following
interpretation: if $f_{ij}$ means the force exerted by the kink $j$ on the kink
$i$ (so that $I_i=f_{i,i+1}+f_{i,i-1}$), then kinematics determines that the
effective force $\tilde f_{i,i\pm 1}$ is a linear combination of 
$(f_{i,i\pm 1}+f_{i\pm 1,i})$ and $f_{i,i\pm 1}$. If $\lambda=0$, the first term
vanishes, but if $\lambda\ne 0$ it does {\it not}: furthermore, for a negative
kink $f_{i\pm 1,i}$ prevails over $f_{i,i\pm 1}$ and it corresponds to a
repulsive force, for kink $i$.

The conclusion we draw from the previous considerations is that 
negative kinks move much slower than positive kinks. 
This results on one side
from the fact that a bigger mass can be attributed to them, and
on the other side that they are subject to an effective repulsive
nn-interaction.

\section{From kink dynamics to coarsening laws}
The interesting dynamical variables are the kink-kink distances $X_i$,
rather than the kink-positions $x_i$. So, from Eq.~(\ref{v_i}) we obtain:
\bea
&&\dot X_i(t) = {a\over 4m_0^2}\times\nonumber\\
&&\left[{1\over X_{i+1}}\left( R^*_\beta(x_{i+2})+R^*_{-\beta}(x_{i+1})
-R^*_\beta(x_{i+1})-R^*_{-\beta}(x_{i})\right)\right.\nonumber\\
&&-\left.{1\over X_{i-1}}\left( R^*_\beta(x_{i})+R^*_{-\beta}(x_{i-1})
-R^*_\beta(x_{i-1})-R^*_{-\beta}(x_{i-2})\right)\right]\nonumber\\
&&+{\sqrt{2F_0}a\over 2m_0}\left[
{\xi_{i+1}\over\sqrt{X_{i+1}}} - {\xi_{i-1}\over\sqrt{X_{i-1}}}\right]
\label{V_i}
\eea

The previous equations have the form:
\bea
\dot q_i(t) &=& {\cal U}_i(\{q\}) + \sum_j {\cal G}_{ij}(\{q\})\xi_j
\hbox{~~~~~with}\\
\langle\xi_j(t)\xi_{j'}(t')\rangle&=&\delta_{jj'}\nonumber
\delta(t-t')
\eea
and we can therefore obtain a Fokker-Planck equation for the
probability $\rho(\{q\},t)$ of finding a given distribution $\{q\}$, 
at time $t$. Two different procedures exist~\cite{Haken}, due to Ito
and to Stratonovich, but as remarked by Kawakatsu and 
Munakata~\cite{KM} the result is the same. This is true even
in presence of the symmetry breaking term, because the two procedures may
differ with respect of the term ${\cal G}_{ij}$, which does not change
if the $\lambda$-term is added.

The Fokker-Planck equation writes:
\be
{\partial\rho\over\partial t} = -\sum_k {\partial\over\partial q_k}
[{\cal U}_k(\{q\})\rho] + {1\over 2}\sum_{k\ell}
{\partial^2\over\partial q_k\partial q_\ell}\sum_m
{\cal G}_{km}{\cal G}_{\ell m}\rho
\ee 

Its actual form, in our case, is~\cite{note3}:
\bea
&&{\partial\rho\over\partial t} = -\sum_k {\partial\over\partial X_k}
[{\cal U}_k(\{X\})\rho]\label{FP}\\ 
&+& {F_0 a^2\over 4m_0^2} \sum_k {1\over X_k}\left[
{\partial^2\over\partial X^2_{k-1}} + {\partial^2\over\partial X^2_{k+1}}
-2 {\partial^2\over\partial X_{k-1}\partial X_{k+1}}\right]\rho\nonumber
\eea
where ${\cal U}_k$ is nothing but the ``deterministic" velocity
of the $k$-th kink.

We are interested in the time dependence of the average value of $X_i$
(which does not depend on $i$). To this end, we define the
distribution functions:
\bea
g(X_i;t) &=& \int_0^\infty (dX)_{\check\imath}\rho\\
g_2(X_i,X_{i+1};t)&=&\int_0^\infty (dX)_{\check\imath ,\check{\imath+1}}\rho\\
g_3(X_i,X_{i+1},X_{i+2};t)&=&\int_0^\infty 
(dX)_{\check\imath ,\check{\imath+1},\check{\imath+2}}\rho
\eea
The notation $(dX)_{\check\imath ,\check{\imath+1},\dots}$ 
means that the integration is
performed on all the variables $X_j$ but $X_i,X_{i+1},\dots$~. 

The details of the calculation follow
Ref.~\cite{KM} and therefore they will not be given here. 
By using the factorization approximation:
\bea
g_2(X_i,X_{i+1};t)&=& g(X_i)g(X_{i+1})\\
g_3(X_i,X_{i+1},X_{i+2};t)&=& g(X_i)g(X_{i+1})g(X_{i+2})
\eea
and integrating Eq.~(\ref{FP}) over $(dX)_{\check\imath}$, we obtain:
\be
{\partial g\over\partial t} = - {\partial\over\partial X} \jmath(X,t)
\label{eq_con}
\ee
with the current of probability given by:
\bea
\jmath(X,t) =&& {a\over 4m_0^2}\left\langle{1\over X}\right\rangle\times
\nonumber\\
&&\left[\langle R^*_+(X)+R^*_{-}(X)\rangle - 
(R^*_+(X)+R^*_{-}(X))\right]g\nonumber\\
&&-{F_0 a^2\over 2m_0^2}\left\langle{1\over X}\right\rangle 
{\partial g\over \partial X}
\eea

In the two relevant limits, $R^*_+(X)+R^*_{-}(X)$ takes the form:
\bea
&&R^*_+(X)+R^*_{-}(X) =\nonumber\\
&& 16\nu m_0^2\exp(-\kappa_0 X)~~~~~\lambda=0
\label{4.4}\\
&&R^*_+(X)+R^*_{-}(X) =\nonumber\\
&& (4\lambda^2 m_0^4/3K) \exp(-\kappa_\lambda X)
~~~\lambda m_0\gg\sqrt{\nu K}\label{4.4new}
\eea
and in the limit $\lambda=0$ we recover Eq.~(4$\cdot$4) of Ref.~\cite{KM}.

Important works on the solution of Eq.~(\ref{eq_con}), which also
go beyond the factorization approximation by
taking into account correlations of consecutive domains, are given 
in a series of papers by Nagai and Kawasaki~\cite{NK}. Here, we will
follow Ref.~\cite{KM} and the first of the papers cited in Ref.~\cite{NK}.

The time dependence of the density of kinks $n(t)$ $-$or
alternatively of the average kink-kink distance: $\overline X(t)\equiv
\langle X\rangle=1/n(t)-$ is studied
by assuming that at large times $\overline X$
represents the only relevant scale in the problem, and therefore $g(X;t)$
satisfies the scaling expression:
\be
g(X;t) = n(t)\tilde g(X/\overline X)
\ee
For example, for a Dirac-delta distribution (all the domains have the same 
size) $\tilde g(s)=\delta (s-1)$, and for a Poisson distribution 
(randomly distributed kinks) $\tilde g(s)=e^{-s}$.

Secondly, we will use a steady-state approximation~\cite{KM} according to which
the distribution $g(X;t)$ does not depend on time, on scales sufficiently
small with respect to $\overline X(t)$: more precisely, on scales
$X<X^*$. This means that the motion of a couple of kinks at distance
smaller than $X^*$ is essentially independent on the position
of all the other kinks. 
Because of the scaling hypothesis, it must result:
$X^*=\overline X/\alpha$, with $\alpha$ constant.

The temporal variation of $n(t)$ is determined by the number of
kink-kink annihilations per unit time and unit length. Since each
annihilation makes two kinks disappear, we have:
\be
\dot n(t) = 2n(t)\jmath(X=0;t) = 2n(t)\jmath(X^*;t)
\label{eq_finale}
\ee
where the second relation derives from the fact that $\partial_t g =0$
implies $\partial_X\jmath=0$.

By approximating $\langle f(X)\rangle$ with $f(\overline X)$, ($f$ is a
generic function), and by neglecting $R^*_\beta(\overline X)$ with
respect to $R^*_\beta(X^*)$, we finally obtain the following expression
for the current in $\overline X^*$:
\bea
\jmath(X^*;t) = &-&{a\over 4m_0^2}{1\over\overline X}
[R^*_+(X^*)+R^*_{-}(X^*)]g(X^*)\nonumber\\
&-&{F_0 a^2\over 2m_0^2}{1\over\overline X}
\left.{\partial g\over\partial X}\right|_{X^*}\nonumber
\eea

\subsubsection{Deterministic regime}\noindent
If the noise term is negligible:
\be
\jmath(X^*;t) = -{a\over 4m_0^2}{1\over\overline X}
[R^*_+(X^*)+R^*_{-}(X^*)]g(X^*)
\ee
Let us consider separately the two limiting cases. When
$\lambda=0$, by using Eq.~(\ref{4.4}), the current writes:
\be
\jmath(X^*;t)=-4a\nu\tilde g(1/\alpha) n^2(t)\exp(-\kappa_0/\alpha n)
\ee
and Eq.~(\ref{eq_finale}) becomes:
\be
\dot n(t) = -8a\nu\tilde g(1/\alpha) n^3(t)\exp(-\kappa_0/\alpha n)
\ee
whose solution gives, at large times:
\bea
&&\overline X(t)\simeq (\alpha/\kappa_0)\ln(t/t_1)\nonumber\\
&&t_1=\left[{e\alpha^2\over 8\tilde g(1/\alpha)}\right]
{K\over a\nu^2}~~~~~~~~(\lambda=0)
\label{log_1}
\eea

In the opposite limit of a strong symmetry breaking ($\lambda m_0\gg
\sqrt{\nu K})$, a similar calculation gives:
\bea
&&\overline X(t)\simeq (\alpha/\kappa_\lambda)\ln(t/t_2)\nonumber\\
&&t_2=\left[{3e\alpha^2\over 4\tilde g(1/\alpha)}\right]
{K\over a\nu^2}~~~~~~~~(\lambda m_0\gg \sqrt{\nu K})
\label{log_2}
\eea

We therefore obtain that $t_1\simeq t_2\simeq t^*$, where $t^*$ was
defined in Sec.~\ref{standard_model} as the time necessary for the
developing of the linear instability of the flat surface. So, the time scale
for the logarithmic coarsening doesn't depend on $\lambda$, but the
length scale does, since it depends on the width of the (largest) domain wall.

We can ask what is the meaning of the $\alpha$-dependence in 
Eqs.~(\ref{log_1}-\ref{log_2}). As pointed out by Nagai and Kawasaki~\cite{NK},
since $\alpha\ln t=\ln t^\alpha$ the parameter $\alpha$ should have some
``universal" value. In a mean-field calculation these authors find
$\alpha=1$, while in a numerical solution of the kink equations they obtain
$\alpha\simeq 3.5$. More rigorous calculations~\cite{NK} give $\alpha=2.27$
if domains are completely uncorrelated, and $\alpha=3.56$ if
correlation effects between neighbouring domains are taken into account.

\subsubsection{Noise-dominated regime}\noindent
Now the current is:
\be
\jmath(X^*;t) = -{F_0 a^2\over 2m_0^2}{1\over\overline X}
\left.{\partial g\over\partial X}\right|_{X^*}
\ee
The equation for $n(t)$ writes:
\be
\dot n(t) = -\left[{F_0 a^2\over m_0^2}\tilde g'(1/\alpha)\right] n^4(t)
\ee
and the solution is:
\bea
&&\overline X(t)=\overline X_0 (t/t_0)^{1/3}\nonumber\\
&&\overline X_0 =\left[{3a\tilde g'(1/\alpha)\over m_0^2}\right]^{1/3}~~~~~~
t_0=1/F_0 a
\eea

So, we will have logarithmic coarsening at ``small" times
and a power-like one at later times. The cross-over time is
determined by the relation $(\alpha/\kappa)\ln(t_c/t^*)=
\overline X_0 (t_c/t_0)^{1/3}$.
By neglecting the logarithmic dependence (also because $t^*\gg t_0$), 
it is found that:
\be
t_c\approx t_0\left({\alpha\over\kappa\overline X_0}\right)^3
\ee

So, the ratio between the cross-over time in presence of a strong asymmetry
and the cross-over time in absence of the $\lambda$-term
is approximately given by:
\be
{t_c(\lambda m_0\gg\sqrt{\nu K})\over t_c(\lambda=0)}
\approx\left({\kappa_0\over\kappa_\lambda}\right)^3=
\left({\lambda m_0\over\sqrt{\nu K}}\right)^3
\label{tc}
\ee

It is important to stress the cubic exponent in the previous expression:
so, even a not large value of $(\kappa_0/\kappa_\lambda)$ gives rise
to a logarithmic coarsening which proceeds for a much longer time,
because kink interaction is stronger and therefore a larger $t_c$ is
necessary so that noise get the better of the deterministic regime.

We want to emphasize that in the noise-dominated regime, the actual value
of $\alpha$ is much less relevant than in the deterministic regime,
because of the power-low character of the coarsening.

\section{Discussion}
\label{Sec_disc}

The main result of the present paper is that ``coarsening laws" don't
change if the symmetry-breaking current $j_{SB}$ is put in the problem
(at least, as far as a continuum local description is valid: see below).
This is mainly due to the fact that the functional form of the kinks does not
change, as shown by the exact solution we have given in
Sec.~\ref{kink_profiles} for their profile.

So, a first question is how general is this result if we modify the
surface current, and therefore Eq.~(\ref{sta_eq}). A first obvious
modification would be to replace $\partial_x A(m^2)=\lambda mm'$ with a more
complicated expression of the slope $m$. This corresponds to have a $\lambda$
depending on $m$; in fact, $\lambda=\lambda(m^2)=2A'(m^2)$.
Since in the late stages of growth the slope is almost everywhere
equal to $\pm m_0$, $\lambda$ is almost everywhere a constant equal to
$\lambda(m_0^2)$.
Is it possible to simply replace $\lambda$ by $\lambda(m_0^2)$ in the final
results? This should not be a bad approximation, as suggested by the
analysis of Eq.~(\ref{sta_eq}) when $\lambda$ depends on $m$.
In fact, the asymptotic behaviour of $M(x)$ [the relevant one for
kink interaction] and the values of $\kappa_\pm$ can be found by
linearizing the differential equation with respect to $(m_0-M(x))$ for a
positive kink and to $(-m_0+M(x))$ for a negative kink [in both cases,
in the limit $x\to\infty$]. Because of the linearization, only the value
$\lambda(m_0^2)$ enters in the problem and therefore determines the profile.

In a similar way, we can take into account a possible $m$-dependence
of the quantity $K$. In this case, such dependence might arise from a
slope-dependent mobility $\Gamma$~\cite{m_Gamma} $-$if $K$ has an equilibrium
origin$-$ or from the dependence on the terrace length $\ell$ of the
probability to nucleate a new terrace~\cite{PV}, if $K$ derives from
nucleation noise.

Let us now discuss the choice of the slope-dependent current:
$j_{ES}=\nu m(1-m^2/m_0^2)$. The only features we require to have
a phase separation process are: $j'_{ES}(m=0)>0$ (to make the flat
surface unstable) and $j_{ES}(m_0)=0$ for some finite value $m_0$
[indeed, $m_0$ must be the first zero of $j_{ES}$]. These features define
the so-called model~I.

Modifications of $j_{ES}$ inside this model do not change the
given picture, as suggested by the analysis of the stationary profile
of the kink (for the sake of semplicity we put $\lambda=0$). If we
linearize the equation:
\be
j_{ES}(m) + Km''(x) = 0
\ee
with respect to $\epsilon(x)=m_0 -m(x)$, we obtain:
\be
j'_{ES}(m_0)\epsilon(x) + K\epsilon''(x) =0
\ee
whose solution is again an exponential function. So, for $x\to\infty$:
$m(x)=m_0 -\epsilon_0 e^{-\kappa x}$, with $\kappa=\sqrt{-j'_{ES}(m_0)/K}$.
In our expression of $j_{ES}$ (Eq.~\ref{current}): $j'_{ES}(m_0)=-2\nu$
and $\kappa$ reduces to $\kappa_0=\sqrt{2\nu/K}$.

Conversely, in model~II there is no finite zero in $j_{ES}$. This implies
that the slope increases with no upper limit: for $\lambda=0$,
as shown by Hunt~et~al.~\cite{Hunt}, the maximal slope $M_0$ in the
profile is asymptotically proportional to the size of the mounds:
$M_0(t)\sim \overline X(t)$. Since the potential energy $U(m)$
[$U'(m)=-j_{ES}(m)$] has no minima, it is no more possible to define
domains and domain-walls, i.e. kinks.

Concerning the time dependence of coarsening, the only
existing numerical results are the ones found by Hunt et al.~\cite{Hunt}.
According to their simulations (in presence of noise),
$\overline X(t)\approx t^n$ with $n\simeq 0.22$\,:
a fairly small value~\cite{small}.
No (rigorous) theoretical derivation of $n$ is available at the moment.
Some scaling arguments $-$applicable to noiseless growth$-$ can be
found in Rost and Krug~\cite{Rost} and in Golubovi\'c~\cite{Golub}:
the former give $n\le 1/4$ while the latter gives the equality
$n=1/4$~\cite{com_Golub}.

A final question we want to face now is how narrow kinks $A$ actually are.
In the limit $\lambda m_0\gg\sqrt{\nu K}$, from Eq.~(\ref{kappa}) we have:
$\kappa_+=\lambda m_0/K$ and $\kappa_-=2\nu/\lambda m_0$. A simple
inspection shows that $[\lambda]=[K]=\hbox{length}^3\cdot
\hbox{time}^{-1}$. Previous evaluations suggest~\cite{PV,Rodi}:
$\lambda\approx K\approx F_0\ell_D^4$. This expression for $K$ is
surely wrong, if thermal detachment plays an important role.
Conversely, if $\lambda$ and $K$ $-$or, more precisely, $\lambda(m_0^2)$
and $K(m_0^2)-$ are of the same order of magnitude, we obtain
$\kappa_+\approx m_0$. This means that
the width of the positive kink $(=1/\kappa_+)$ is
nothing but the inverse of the value of the constant slope in the
surface profile: so, if $m_0$ is determined by the symmetry of the
crystal lattice, $m_0\simeq 1/a$ and the positive kink is as narrow as
a lattice constant! In this case, our description would break down,
because the regions of positive curvature in the surface profile
would correspond to a discontinuity of the slope, i.e. to angular points,
which are not compatible with a local continuum equation~\cite{notePV}.

\section{Conclusions}

The kink picture not only has allowed to find the coarsening law in
presence of the symmetry breaking term, but it has also given a
qualitative description of the dynamics which allows a better
comprehension of the evolution of the system:
the widening or the narrowing of a kink; the consequent different
velocities of kinks $A$ and $B$; the conservation of the order
parameter seen as a kinematical constraint on kink movement;
the difference between the ``real" force acting on a kink and the
``effective" force felt by the kink, because of such constraint.

In this respect, the most important consequence of the breaking of
symmetry is that negative kinks feel an effective {\it repulsive}
interaction with the nn kinks (but attractive with the nnn ones).
It is important to stress this point because
coarsening is the result of a global
attraction between kinks: if kinks repelled each other, the
configuration with the $X_i$ all equal would be stable.

Finally, the kink picture has provided the condition of applicability
of the local theory:
\be
{1\over\kappa_+} = {K(m_0^2)\over m_0\lambda(m_0^2)} \gg a
\ee
If this relation is not fulfilled, a different method to study coarsening
should be used. In Ref.~\cite{PV} we showed that in this case the
evolution of the surface is governed by a nonlocal current;
alternatively, we can keep a local description, but we must add a
singular term to the current $j$, and couple the Langevin equation:
$\partial_t z(x,t) = -a\partial_x j$ with specific evolution equations
for the angular points. It would be clearly interesting to check if
a different coarsening process may arise from an ``angular point" picture.

To our knowledge, the current (\ref{current}) has not been
formerly studied. The closest model is the one considered by
Stroscio et al.~\cite{Stroscio} in two dimensions, where the Mullins
term $(Km''(x))$ is replaced by a higher order one ($Km''''(x)$)
and the resulting equation is studied numerically.
Clearly, in two dimensions analytical treatments are much more difficult;
anyway, neither a numerical solution of the model studied in the present
paper is available at the moment.
One reason is that in two dimensions, even the model without $j_{SB}$ is
not yet fully understood, since the evolution equation for $m(x,t)$%
~\cite{Siegert} is no more equivalent to the Cahn-Hilliard equation.

\acknowledgments
I warmly thank Joachim Krug for several useful discussions and for a
very careful reading of the manuscript. I also gratefully
acknowledge the Alexander von Humboldt Stiftung for financial support.

\appendix
\section{Langevin equations for the kinks}
\label{app_A}

\subsection{Absence of noise}
The starting point is the following multi-kink expansion:
\bea
m(x,t) &=& M_i(x,t) + \sum_{j>i} [M_j(x,t)-M_j(-\infty)]\nonumber\\
                    &+& \sum_{j<i} [M_j(x,t)-M_j(\infty)]\\
       &\equiv& M_i(x,t) + \delta m_i 
\eea
which gives rise, once replaced in (\ref{1}), to:
\be
\am\sum_j \left[-v_j M'_j +\dot v_j {\partial M_j\over\partial v_j}\right]
= \D j[m(x,t)]
\label{4}
\ee

$M_j$ depends on $x$ and $t$ through the combination $(x-v_j t)$ and
$M'_j$ is the derivation with respect to all this argument. We will
also use the notation: $\d M$ to mean the same kind of derivation. 
The single kink profile is found by simply dropping the sum $\sum_j$
and the term in $\dot v$ in Eq.~(\ref{4}):
\be
-\am v^0_j M'_j = \D j[M_j]
\label{5}
\ee

It will be useful to consider, together with $M_j$, also its spatial
derivative $M'_j$ which is localized around $x=x_j$. We define also
$\tilde M'_j$ through the relation: $M'_j(x) = \D \tilde M'_j (-x)$.
They satisfy the relations:
\bea
&&-\am v^0_j M'_j(x) =\nonumber\\
&& \D [K\d^2 -U''(M_j)+\lambda M'_j +\lambda M_j\d]M'_j(x)
\label{6}\\
&&\am v^0_j \tilde M'_j(x) =\nonumber\\
&& [K\d^2 -U''(M_j)+\lambda M'_j +\lambda M_j\d]
\D\tilde M'_j(x)\label{7}
\eea

Now, let us multiply (\ref{4}) by $\tilde M'_i(x)$ and integrate on $x$.
By defining $\delta v_j\equiv v_j -v^0_j$, we can write:
\bea
&&\am\sum_j \int dx\tilde M'_i(x)\left\{ -\delta v_j M'_j +
\dot v_j{\partial M_j\over\partial v_j}\right\} =\nonumber\\ 
&&\int dx \tilde M'_i \D j[m] + \am\sum_j v^0_j \int dx \tilde M'_i M'_j
\label{8}
\eea

The next step is to replace $m(x,t)=M_i(x)+\delta m_i$ in the current
$j[m]$. The definition of the nonlinear part of the potential $U(m)$ 
(or equivalently of the current $U'(m)$) is self-explanatory.
\bea
j[m] &=& Km''-U'(m)+\lambda mm'\nonumber\\
&=& KM''_i +K\delta m''_i - U'(M_i+\delta m_i)\nonumber\\
&&+\lambda(M_i+\delta m_i) (M'_i+\delta m'_i)\nonumber\\
&=& KM''_i +K\delta m''_i - U'(M_i) -U''(M_i)\delta m_i
-U'_{NL,i}\nonumber\\
&& +\lambda M_i M'_i +\lambda\delta m_i M'_i +\lambda M_i\delta m'_i
+\lambda\delta m_i\delta m'_i\label{10}
\eea

The three terms which do not depend on $\delta m_i$, once used Eq.~(\ref{5})
cancel the term $j=i$ in the last summation of Eq.~(\ref{8});
$\delta m''_i$ is simply written as $\sum_{j\ne i} M''_j$, whilst all the
other terms for the moment keep unchanged. So, the right hand side of
Eq.~(\ref{8}) rewrites:
\bea
&&RHS|_{(\protect\ref{8})} =\nonumber\\
&&\int dx \tilde M'_i(x)\left\{
\sum_{j\ne i} [K\D M''_j +\am v^0_j M'_j]\right.\nonumber\\ 
&&- \D[U'_{NL,i}+U''(M_i)\delta m_i] \nonumber\\
&&+\lambda\D[\delta m_i M'_i
+M_i\delta m'_i +\delta m_i\delta m'_i]\}
\label{11}
\eea

Let us consider separately some terms:
$$
\int dx \tilde M'_i(x)\sum_{j\ne i}K \D M''_j =
\int dx K\delta m_i \d^2 \D \tilde M'_i
$$
$$
-\int dx \tilde M'_i(x) \D [U''(M_i)\delta m_i] =
-\int dx U''(M_i)\delta m_i \D\tilde M'_i
$$
\bea
&&\lambda\int dx \tilde M'_i\D[\delta m_i M'_i +\delta m'_i M_i] =\nonumber\\
&&\lambda\int dx [\delta m_i M'_i +2M_i \delta m'_i]\D \tilde M'_i\nonumber\\
&&+\lambda\int dx\delta m_i [M'_i +M_i\d]\D\tilde M'_i\nonumber
\eea

Eq.~(\ref{11}) therefore rewrites:
\bea
&&RHS|_{(\protect\ref{8})} =\nonumber\\ 
&&\int dx\delta m_i [K\d^2-U''(M_i)+\lambda
M'_i +\lambda M_i\d]\D\tilde M'_i\nonumber\\
&&+\int dx\tilde M'_i\sum_{j\ne i}\am v^0_j M'_i\nonumber\\
&&+\int dx\tilde M'_i\left[-\D U'_{NL,i}+\lambda\D (\delta m_i M'_i +
2M_i\delta m' +\delta m_i\delta m'_i)\right]\nonumber\\
&&\equiv (A) + (B) + (C)
\eea
This way, Eq.~(\ref{8}) takes the form ($LHS\equiv$left-hand-side):
$LHS|_{(\protect\ref{8})}=(A)+(B)+(C)$.

By using Eq.~(\ref{7}):
\be
(A)=\am \int dx\tilde M'_i\sum_{j\ne i}[-v_i^0 M'_i]
\ee
which can be  summed to $(B)$, giving:
\be
(A)+(B)=\am \int dx \tilde M'_i\sum_j (v_j^0-v_i^0)M'_j
\ee
and subtracting $LHS|_{(\protect\ref{8})}$:
$$
(A)+(B)-LHS|_{(\protect\ref{8})}=-(C)
$$
that is to say:
\bea
&&\am\sum_j\left[(v_j-v_i^0)
(\tilde M'_i,M'_j)-\dot v_j(\tilde M'_i,{\partial M_j\over\partial v_j})
\right]\nonumber\\
&=& \int dx\tilde M'_i[\D U'_{NL,i}\nonumber\\
&&-\lambda\D
(\delta m_i M'_i + 2M_i\delta m' +\delta m_i\delta m'_i)]
\label{13}\\
&\equiv & (C_1)+(C_2)+(C_3)+(C_4)\nonumber
\eea
In the previous equation we have used the following scalar product:
\be
(R,S)=\int_{-\infty}^{+\infty} dx R(x)S(x)
\ee
The three terms in square brackets on the right-hand-side 
$[(C_2)+(C_3)+(C_4)]$ represent the effect of the symmetry breaking current.

By integrating by parts and by using the definition of $\tilde M'_i$:
\be
(C_1)=(M'_i(-x),U'_{NL,i})
\ee
If we define the function $G(x,y)\equiv U'(x+y)-U'(x)-yU''(x)$, then
$U'_{NL,i}=G(M_i,\delta m_i)$. 
In the following, we will also make use of the function:
$\tilde U(x,y)\equiv U(x+y)-U(x)-yU'(x)$. It is obvious that
$G(x,y)=\partial_x\tilde U(x,y)$.

We observe that: \i)~$G(M_i,0)=0$;
\i\i)~$G$ may be written as a Taylor expansion whose generic term
contains $(\delta m_i)^n$; \i\i\i)~$G$ is not linear in $\delta m_i$,
but if we use the binary interaction approximation, it is indeed linear.
This approximation corresponds to write:
$$
(\delta m_i)^n\approx\sum_{j>i} [M_j-M_j(-\infty)]^n
+\sum_{j<i} [M_j-M_j(\infty)]^n
$$
In this approximation, we obtain:
\bea
(C_1)=\sum_{j>i}\int dx M'_i(x_i-x)G(M_i(x-x_i),M_j-M_j(-\infty))&&\nonumber\\
+ \sum_{j<i}\int dx M'_i(x_i-x)G(M_i(x-x_i),M_j-M_j(\infty))&&\nonumber
\eea

We must observe that $M'_i$ is not vanishing only when $x\approx x_i$;
furthermore, $[M_j-M_j(\pm\infty)]$ goes to zero when $(+)$ $x>x_j$
or $(-)$ $x<x_j$. On the basis of these considerations, it is possible
to write:
\bea
(C_1) = \sum_{j>i}\int_{x_j}^{+\infty} dx M'_i(x_i-x)G(M_i(x-x_i),\Delta M_j)
&&\nonumber\\
+ \sum_{j<i}\int_{-\infty}^{x_j} dx M'_i(x_i-x)G(M_i(x-x_i),-\Delta M_j)
&&\nonumber
\eea
where $\Delta M_j\equiv M_j(\infty)-M_j(-\infty)$.

Since $M'_i(x)$ is an even function of $x$:
\bea
&&\int_{x_a}^{x_b} dx M'_i(x_i-x)G(M_i(x-x_i),\hbox{const}) =\nonumber\\
&&\left.\tilde U(M_i(x-x_i),\hbox{const})\right|^{x_b}_{x_a}\nonumber
\eea
and $(C_1)$ can be written as:
\bea
(C_1) = \sum_{j>i}[\tilde U(M_i(\infty),\Delta M_j)-
\tilde U(M_i(x_j-x_i),\Delta M_j)]&&\nonumber\\
+ \sum_{j<i}[\tilde U(M_i(x_j-x_i),-
\Delta M_j)-\tilde U(M_i(-\infty),-\Delta M_j)]&&\nonumber
\eea
At the first order in the small quantities $[M_i(x_j-x_i)-M_i(\pm\infty)]$
($\pm$ resp. for $j>i$ and $j<i$), we have:
\bea
&&(C_1)=-\sum_{j>i}[M_i(x_j-x_i)-M_i(\infty)]G(M_i(\infty),\Delta M_j)
\nonumber\\
&&+\sum_{j<i}[M_i(x_j-x_i)-M_i(-\infty)]G(M_i(-\infty),-\Delta M_j)
\label{14}
\eea

In the following, we will restrict ourselves to nearest-neighbour kinks
interaction, and therefore only the terms $j=i\pm 1$ will survive
in (\ref{14}). If we also use the fact that:
\be
G(M_i(\pm\infty),\pm\Delta M_{i\pm 1}) =
\mp\Delta M_{1\pm 1}U''(M_i(\pm\infty))
\ee
we obtain the following final expression:
\bea
&&(C_1)=[M_i(x_{i+1}-x_i)-M_i(\infty)]\Delta M_{i+1} U''(M_i(\infty)) 
\nonumber\\
&&+ [M_i(x_{i-1}-x_i)-M_i(-\infty)]\Delta M_{i-1} U''(M_i(-\infty))
\label{15}
\eea

The procedure to follow for the treatment of the other terms $(C_i)$ is
similar. In poor words, if $R(x)$ and $S(x)$ are functions which are
localized, resp. in $x_1$ and $x_2$, we make the approximation:
\bea
R(x-x_1)S(x-x_2)\approx&& R(x-x_1)S(x_1-x_2)\nonumber\\
&+& R(x_2-x_1)S(x-x_2)\nonumber
\eea
and then we retain only the term corresponding to the function
decreasing more rapidly (for example, if $R(x)$ was a Dirac-delta, only
the first term would be retained, because the second one would be exactly zero).
We give here only the results.
\bea
(C_2)&=&\lambda\{[M_{i+1}(x_{i+1}-x_i)-M_{i+1}(\infty)]\nonumber\\ 
&-&[M_{i-1}(x_{i}-x_{i-1})-M_{i-1}(\infty)]\}\int_{-\infty}^{+\infty}
dx (M'_i)^2\nonumber\\
(C_3)&=&0\nonumber\\
(C_4)&=&-{\lambda\over 2}(\Delta M_{i+1})^2 M'_i(x_{i+1}-x_i)\nonumber\\
&+&{\lambda\over 2}(\Delta M_{i-1})^2 M'_i(x_{i}-x_{i-1})\nonumber
\eea
The expression $(C_3)=0$ means that such term is always of higher order
than the others.

From now on, notation must take into account explicitly the existence of two
different classes of kinks. By using the following results:
\bea
&&M_\beta(x)-M_\beta(\infty)\simeq-\beta 2m_0\exp(-\kappa_\beta x)
~~~~~\hbox{when~~}x\to\infty\nonumber\\
&&R_\beta(x)\equiv\exp(-\kappa_\beta x)\nonumber\\
&&\Delta M_i=\beta 2m_0\nonumber\\
&&U''(\pm m_0)=2\nu\nonumber\\
&&\int_{-\infty}^{+\infty} dx[M'_\beta(x)]^2={2\over 3}m_0^2\kappa_\beta
\nonumber
\eea
it is straightforward to write:
\bea
(C_1) &=& 8\nu m_0^2 [R_\beta(X_i)-R_\beta(X_{i-1})]\nonumber\\
(C_2) &=& \beta (4/3)m_0^3\kappa_\beta\lambda 
[R_{-\beta}(X_i)-R_{-\beta}(X_{i-1})]\label{Ci_A}\\
(C_4) &=& -\beta 4m_0^3\kappa_\beta\lambda [R_\beta(X_i)-R_\beta(X_{i-1})]
\nonumber
\eea

We can now put together Eq.~(\ref{Ci_A}) with the previous ones. A further
approximation is to neglect the ``deformation" of the kink-profile, due
to its velocity, and to suppose that kinks are immobile in absence of
interactions. This way, we obtain:
\be
\am\sum_j v_j (\tilde M'_i,M'_j) = (C_1)+(C_2)+(C_4)
\ee
where the $LHS$ can be further developed:
\bea 
&&\am\sum_j v_j (\tilde M'_i,M'_j)=\nonumber\\
&&\am\sum_j v_j \int dx \tilde M'_i(x)M'_j(x)=\nonumber\\
&&\am\sum_j v_j \int dx\D\tilde M'_i(x)\D^{-1}M'_j(x)=\nonumber\\
&&\am\sum_j v_j \int dx M'_i(x)\D^{-1}M'_j(x)
\label{E}
\eea

We have therefore to determine the inverse of the operator $\D$. By
following Kawasaki and Ohta~\cite{KO}:
\be
\D^{-1} A(x,t) = -{1\over 2}\int dx' |x-x'| A(x',t)
\label{D-1}
\ee
The ``integration constants" appearing when the operator $\D$ is inverted,
are shown to be irrelevant for the kink dynamics (Ref.~\cite{KM}).

By applying Eq.~(\ref{D-1}) to Eq.~(\ref{E}):
\bea
&&\int dx M'_i(x)\D^{-1}M'_j(x) =\nonumber\\
&& -{1\over 2}\int\int dx dx' M'_i(x)|x-x'|M'_j(x')\\
&\approx&-{1\over 2}|x_i-x_j|\int\int dx dx'M'_i(x-x_i)M'_j(x-x_j)\\
&=&-{1\over 2}|x_i-x_j|\Delta M_i\Delta M_j
\eea
So, Eq.~(\ref{13}) finally writes:
\be
-\am{\Delta M_i\over 2}\sum_j\Delta M_j |x_i-x_j|\dot x_j = (C_1)+(C_2)+(C_4)
\label{pinocchio_A}
\ee

\subsection{The effect of noise}

The term of noise $\delta F(x,t)$ in Eq.~(\ref{lan_eq}) corresponds to a term
$\eta(x,t)=\partial_x\delta F(x,t)$ on the right-hand-side of 
Eq.~(\ref{4}). To see how it affects the kink movement, it must be
multiplied by $\tilde M'_i(x)$ and integrated on $x$. Since the $LHS$ of
(\ref{pinocchio_A}) indeed corresponds to minus the $LHS$ of (\ref{4}),
if we call $\eta_i(t)$ the noise term to be added to $(C_1)+(C_2)+(C_4)$
in (\ref{pinocchio_A}), it will result:
\be
\eta_i(t)=-\int dx \tilde M'_i(x)\eta(x,t)=\int dx\partial_x \tilde M'_i(x)
\cdot\delta F(x,t)
\ee
The following properties are found~\cite{KM}:
$$
\langle\eta_i(t)\rangle=0~~~~~~~\hbox{and}
$$
\bea
\langle\eta_i(t)\eta_j(t')\rangle&=&2F_0\delta(t-t')\int dx
\tilde M'(x)\D\tilde M'_j(x)\nonumber\\
&=&-4m_0^2 F_0(-1)^{i-j}|x_i-x_j|\delta(t-t')
\eea
To derive the spatial correlation between noise, we have used the
definition of $\tilde M'_i$ and inverted the operator $\D$. Finally, we
have used the fact that $\Delta M_i\Delta M_j = 4m_0^2 (-1)^{i-j}$, a
relation which can be used also for the $LHS$ of Eq.~(\ref{pinocchio_A}).
So, we obtain the following system of coupled Langevin equations:
\bea
-2\am m_0^2\sum_j (-1)^{i-j}|x_i-x_j|\dot x_j =&&\nonumber\\
 (C_1)+(C_2)+(C_4) + \eta_i(t)&&
\eea


\begin{references}
\item[$^a$]
E-mail address: {\tt politip@fi.infn.it}

\bibitem{JVAP}
J. Villain and A. Pimpinelli, Physique de la Croissance
Cristalline, Eyrolles-Alea-Saclay, Paris (1994) and english
version (Cambridge University Press), in press.

\bibitem{book_BS}
A.L.Barab\'asi and H.E. Stanley, Fractals concepts in surface growth,
Cambridge University Press, Cambridge (1995). In particular, see
Chapter~27.

\bibitem{noteW}
First studies of the nucleation noise for a one-dimensional surface,
are: D. Wolf, in ``Scale invariance, interfaces, and non-equilibrium
dynamics", A. McKane et al. eds. (Plenum Press, New~York, 1995);
E.~Somfai, D.E.~Wolf and J.~Kert\'esz, J. de Physique I {\bf 6},
393 (1996).

\bibitem{ES}
A recent discussion is:
K. Kyuno and G. Ehrlich, Surf.~Science {\bf 383}, L766 (1997).

\bibitem{JV}
J. Villain, J. de Physique I {\bf 1}, 19 (1991).

\bibitem{PV}
P. Politi and J. Villain, Phys. Rev. B {\bf 54}, 5114 (1996).

\bibitem{z_inv}
We leave out the possibility of a current that is periodic in $z$
($j=j_0\sin(2\pi z)$), because it would be an effect of the discrete
nature of the lattice, which may be important only in the very early
stages of growth.

\bibitem{P}
P. Politi, J. Phys. I France {\bf 7}, 797 (1997).

\bibitem{review_Krug}
J. Krug, Adv. Phys. {\bf 46}, 139 (1997).

\bibitem{KPS} J. Krug, M. Plischke and M. Siegert, Phys. Rev. Lett.
{\bf 70}, 3271 (1993);
M. Siegert and M. Plischke, Phys. Rev. Lett. {\bf 73}, 1517 (1994).

\bibitem{AF}
J.G. Amar and F. Family, Surf. Science {\bf 365}, 177 (1996).

\bibitem{non_thermal}
J.W. Evans, Phys. Rev. B {\bf 43}, 3897 (1991).

\bibitem{SV}
P. \v{S}milauer and D.D. Vvedensky, Phys. Rev. B {\bf 52}, 14263 (1995).

\bibitem{note1}
Here $j_{ES}$ must be read as the ``slope-dependent current", i.e. the
current due to the Ehrlich-Schwoebel effect, and also to all possible
different mechanisms which depend on $m$ (see above).
Current (\protect{\ref{j_sI}}) has been used in the context of surface growth
by Stroscio et al.~\protect\cite{Stroscio}.
The fact that it diverges when $|m|\to\infty$ is not relevant, because
$m=m_0$ is a ``stable fixed point".
Current (\protect{\ref{j_sII}}), which is correct only in the limit of
a strong Ehrlich-Schwoebel effect (see Eq.~(\protect{\ref{j_s2}})),
has been introduced in 2+1 dimensions by:
M.D.~Johnson, C.~Orme, A.W.~Hunt, D.~Graff, J.~Sudijono, L.M.~Sander and
B.G.~Orr, Phys. Rev. Lett. {\bf 72}, 116 (1994).

\bibitem{Mullins}
W.W. Mullins, J. Appl. Phys. {\bf 28}, 333 (1957).

\bibitem{Rodi}
P. Politi and J. Villain, in ``Surface diffusion: atomistic and
collective processes". Eds. M. Scheffler and M. Tringides
(Plenum Press, 1997).

\bibitem{SP} J.A. Stroscio and D.T. Pierce,
Phys. Rev. B {\bf 49}, 8522 (1994).

\bibitem{Stroscio}
J.A. Stroscio D.T. Pierce, M.D. Stiles, A. Zangwill and
L.M. Sander, Phys. Rev. Lett. {\bf 75}, 4246 (1995).

\bibitem{Sun}
T. Sun, H. Guo and M. Grant, Phys. Rev. A {\bf 40}, 6763 (1989).

\bibitem{KPZ} M. Kardar, G. Parisi and Y.-C. Zhang,
Phys. Rev. Lett. {\bf 56}, 889 (1986).

\bibitem{SS}
E. Somfai and L.M. Sander, in ``Dynamics of crystal surfaces and interfaces".
Eds. P.M.~Duxbury and T.J.~Pence (Plenum Press, 1997).

\bibitem{krug2} J. Krug, in ``Dynamics of fluctuating interfaces and
related phenomena". Eds.~D. Kim et al.  (World Scientific, Singapore 1997).

\bibitem{Hunt}
A.W. Hunt, C. Orme, D.R.M. Williams, B.G. Orr and L.M. Sander,
Europhys. Lett. {\bf 27}, 611 (1994); and
in ``Scale invariance, interfaces, and non-equilibrium
dynamics", A. McKane et al. eds. (Plenum Press, New~York, 1995).
In these papers authors study the model with $\lambda=0$ and
an unstable current $j_{ES}$ as
given by the model~II. They draw information on the dynamics from the
numerical evaluation of the smallest eigenvalue of a proper operator
associated to the Langevin equation: anyway, no analytical evaluation
is given.

\bibitem{opl}
In a different context [O. Pierre-Louis, C.~Misbah, Y.~Saito,
J.~Krug and P.~Politi, Phys. Rev. Lett. {\bf 80}, scheduled for the issue
of 4~May 1998] it is indeed possible to observe no coarsening because
the period is a decreasing function of the amplitude.

\bibitem{review_Bray}
A.J. Bray, Adv. Phys. {\bf 43}, 357 (1994).

\bibitem{KO}
K. Kawasaki and T. Ohta, Physica {\bf 116A}, 573 (1982).

\bibitem{KM}
T. Kawakatsu and T. Munakata, Prog. Theo. Phys. {\bf 74}, 11 (1985).

\bibitem{Haken}
H. Haken, Advanced synergetics (Springer, Berlin, 1983).


\bibitem{note3}
Eq.~(4$\cdot$1) of Ref.~\protect\cite{KM} contains indeed a misprinting:
$\partial^2/\partial X_j^2$ should be replaced by
$\partial^2/\partial X_{j-1}\partial X_{j+1}$.

\bibitem{NK}
T. Nagai and K. Kawasaki, Physica {\bf 120A}, 587 (1983);
K. Kawasaki and T. Nagai, Physica {\bf 121A}, 175 (1983);
T. Nagai and K. Kawasaki, Physica {\bf 134A}, 483 (1986).

\bibitem{m_Gamma}
J. Krug, H.T. Dobbs and S. Majaniemi, Z. Phys. B {\bf 97}, 281 (1995).

\bibitem{small}
For model I, $n=1/3$ for a conserved order parameter and $n=1/2$
for a nonconserved one.

\bibitem{Rost}
M. Rost and J. Krug, Phys. Rev. E {\bf 55}, 3952 (1997).

\bibitem{Golub}
L. Golubovi\'c, Phys. Rev. Lett. {\bf 78}, 90 (1997).

\bibitem{com_Golub}
This value is a bit surprising: if compared to $n\simeq 0.22$ it would
lead to conclude that (in 1+1 dimensions) deterministic coarsening is not
slower than the noisy one; if compared to the noiseless coarsening of
model~I ($L(t)\sim\ln t$), we should conclude that steepening (due to the
absence of finite zeros in $j_{ES}$) {\it favours} the coarsening.

\bibitem{notePV}
The reason is simply that $j_{SB}$ (the cause of angular points)
would contribute to the growth velocity
with a term proportional to $\partial_x^2 A(m^2)$, which diverges in the
angular points. More details are given in Ref.~\protect\cite{PV}.

\bibitem{Siegert}
M. Siegert, Physica A {\bf 239}, 420 (1997).


\end{references}
\end{document}